\documentclass[preprint,12pt]{elsarticle}
\usepackage{amssymb}

\usepackage{lineno}
\usepackage{color}

\journal{Nuclear Physics B}

\begin{document}

\begin{frontmatter}



\title{An overview of the Daya Bay Reactor Neutrino Experiment}


\author[ihep]{Jun Cao}
\ead{caoj@ihep.ac.cn}

\author[lbnl]{Kam-Biu Luk}
\ead{k{\_}luk@berkeley.edu}

\address[ihep]{Institute~of~High~Energy~Physics, Beijing, China}
\address[lbnl]{Department of Physics, University~of~California and Lawrence~Berkeley~National~Laboratory, Berkeley, California, USA}

\begin{abstract}
The Daya Bay Reactor Neutrino Experiment discovered an unexpectedly large neutrino oscillation related to the mixing angle $\theta_{13}$ in 2012. This finding paved the way to the next generation of neutrino oscillation experiments. In this article, we review the history, featured design, and scientific results of Daya Bay. Prospects of the experiment are also described.
\end{abstract}


\end{frontmatter}


\section{Introduction}
\label{introduction}

\par
Neutrino oscillation was firmly established by 2002. Around that time, atmospheric and accelerator neutrino experiments, e.g. Super-K~\cite{superk} and K2K~\cite{k2k}, have determined the oscillation parameters $\theta_{23}$ and $|\Delta m^2_{32}|$ whereas solar and reactor neutrino experiments, such as SNO~\cite{sno} and KamLAND~\cite{kamland}, measured $\theta_{12}$ and $\Delta m^2_{21}$. However, the mixing angle $\theta_{13}$, the CP violating phase $\delta_{\rm CP}$, and the sign of $\Delta m^2_{32}$ (aka the mass hierarchy) were unknown. In addition, $\theta_{13}$, unlike the other two mixing angles, was expected to be small~\cite{chooz,paloverde}.
\par
Among the three unknown quantities, $\theta_{13}$ plays a critical role in defining the future experimental program on neutrino oscillation. It is known that the CP-violating effect is proportional to
\begin{equation}
J = \sin 2\theta_{12} \sin 2\theta_{23} \sin 2\theta_{13} \cos\theta_{13} \sin\delta_{\rm CP} \approx 0.9 \sin2\theta_{13} \sin\delta_{\rm CP} \,.
\end{equation}
Resolution of the mass hierarchy problem also relies on the size of $\theta_{13}$. If it is too small, current technologies may not be able to determine $\delta_{\rm CP}$ and the mass hierarchy.

\par
The mixing angle $\theta_{13}$ can be measured by accelerator-based or reactor-based experiments. However, the appearance probability of $\nu_{\mu} \to \nu_e$ in an accelerator neutrino experiment also depends on the yet unknown $\delta_{\rm CP}$ and the mass hierarchy. Hence, this type of experiments can only provide evidence for a non-zero $\theta_{13}$ but cannot measure its value unambiguously. On the other hand, reactor-based experiments
can unambiguously determine $\theta_{13}$ via measuring the survival probability of the electron antineutrino $\overline{\nu}_{e}$ at short distance (${\cal O}(km)$) from the reactors. In the three-neutrino framework, the
survival probability is given by
\begin{equation}\label{eq:psurv}
P = 1 - \cos^4\theta_{13}\sin^2 2\theta_{12}\sin^2\Delta_{21} - \sin^2 2\theta_{13}\sin^2
\Delta_{ee}\,,
\end{equation}
\noindent where $\sin^2\Delta_{ee} \equiv \cos^2\theta_{12}\sin^2
\Delta_{31}+\sin^2\theta_{12}\sin^2{\Delta_{32}}\,$
and
$\Delta_{ji}\equiv 1.267 {\Delta}m^2_{ji} L/E$. ${\Delta}m^2_{ji}$  is the mass-squared
difference in eV$^2$, $E$ is the energy of the $\overline{\nu}_{e}$ in MeV, and $L$ is the distance in meters from the production point.

\par
Pinning down $\theta_{13}$ by performing a relative measurement with a set of near and far detectors was suggested at the beginning of this millennium~\cite{mikaelyan}. This method allows cancellation of most systematic uncertainties due to the reactor and the detector that previous experiments suffered. Since
2002, eight reactor experiments were proposed~\cite{whitepaper}; three of them, Daya Bay~\cite{dybcdr}, Double Chooz~\cite{doublechooz}, and RENO~\cite{reno}, were finally constructed.

\par
Among the eight proposals, the Daya Bay experiment is most sensitive for measuring $\theta_{13}$. The nuclear-power  complex is among the top five most powerful in the world, providing a very intense flux of antineutrinos. In addition, it is very close to a mountain range in which an array of horizontal tunnels can be built, providing sufficient
overburden to attenuate cosmic rays and space to accommodate a relatively large-scale experiment.

\par
The Daya Bay nuclear-power complex is located on the southern coast of China, 55 km to the northeast of Hong Kong and 45 km to the east of Shenzhen.  As shown in Fig.~1, the nuclear complex consists of six reactors grouped into three pairs with each pair referred to as a nuclear power plant (NPP)\@.   All six cores are functionally identical pressurized water reactors, each with a maximum of 2.9 GW of thermal power. The last core started commercial operation on 7 August 2011, a week before the start-up of the Daya Bay experiment. The distance between the cores for each pair is 88 m. The Daya Bay cores are separated from the Ling Ao cores by about 1100 m, while the Ling Ao-II cores are around 500 m away from the Ling Ao cores.

\begin{figure}[!htb]
\begin{center}
\includegraphics[width=0.6\textwidth]{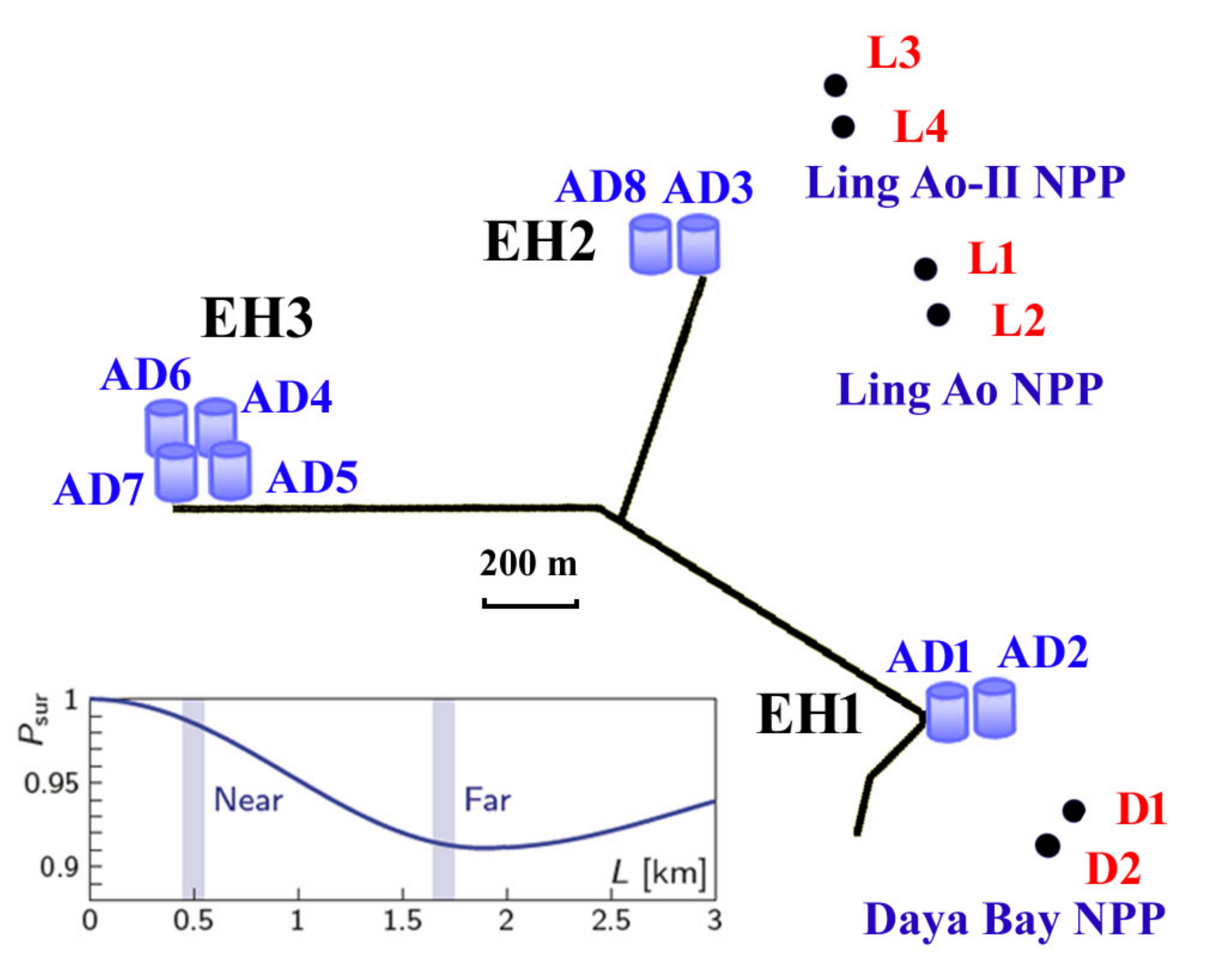}
\caption{Layout of the Daya Bay experiment.
The dots represent the reactor cores, labelled as D1, D2, L1 to L4.
Eight antineutrino detectors, labelled as AD1 to AD8, are installed in three underground experimental halls (EH1-EH3). The bottom sub-panel shows the survival probability as a function of the effective baseline $L$. The near and far detectors locate in the shaded area. \label{fig:layout}}
\end{center}
\end{figure}

\par
The Daya Bay experiment consists of three underground experimental halls (EHs) connected with horizontal tunnels. The overburden for the Daya Bay near hall (EH1), the Ling Ao near hall (EH2) and the far hall (EH3) are 250, 265, and 860 equivalent meters of water, respectively. Eight antineutrino detectors (ADs) are installed in the three halls, with two in EH1, two in EH2, and four in EH3. Each AD has 20-ton target mass to catch the reactor antineutrinos. The sensitivity to $\sin^22\theta_{13}$ was designed to be better than 0.01 at 90\% confidence level in 3 years.

\section{History of the Daya Bay Experiment}
\label{history}

\par
The idea of determining $\theta_{13}$ using the Daya Bay reactor complex was proposed in 2003.
The first dedicated workshop for the Daya Bay experiment was held in the University of Hong Kong in November 2003~\cite{hk2003}. It was immediately followed by the second one in January 2004 at the Institute of High Energy Physics~\cite{ihep2004}, at which a preliminary experimental design was presented, including the unique multiple-detector scheme and the reflective panel design for light collection. In response to the recommendation of measuring $\sin^22\theta_{13}$ to the level of 0.01 by the APS Neutrino Study~\cite{numatrix} and NuSAG~\cite{nusag}, the target mass of the detectors was enlarged from 8 ton to 20 ton. The Conceptual Design Report (CDR) was released at the end of 2006~\cite{dybcdr}.

\par
The Daya Bay Collaboration with international participation was formed in February 2006. The project was approved by the Chinese Academy of Sciences in May 2006 and by the Ministry of Science and Technology of China in January 2007. It passed the CD-2 review as required by the US Department of Energy in January 2008.

\par
Ground breaking of the experiment took place in October 2007. Detectors were assembled onsite in parallel to the civil construction. After 4 years of construction, the first of the three underground experimental halls, EH1, started data taking on August 11, 2011. Data was used to study the detector performance, resulting a paper submitted on 28 February 2012~\cite{ad12}. This publication reported that the relative detection uncertainty of two ADs was only 0.2\%, much better than the designed value of 0.38\% documented in the CDR.

\par
Since the detector fabrication was out of sync with the civil construction, the collaboration decided to operate the experiment with two phases to maximize the scientific reach. The first phase was run with 6 ADs out of a total of 8, with two in EH1, one in EH2, and three in EH3. The second near hall, EH2, was ready on 5 November 2011, and EH3 started data taking on 24 December 2011. On 8 March 2012, the Daya Bay collaboration announced the discovery of a new disappearance of reactor antineutrinos at 5.2 standard deviations ($\,\sigma$) and measured $\sin^22\theta_{13}=0.092\pm0.016{\rm (stat)}\pm0.005{\rm (syst)}$ with 55 days data~\cite{DB_discovery}. It was further verified at 7.7$\,\sigma$ with 139 days of data~\cite{DB_CPC} and at 4.6$\,\sigma$ with a statistically independent data set using antineutrino events tagged by neutron capture on hydrogen~\cite{DB_nH}.

\par
The six-detector phase terminated on 28 July 2012. The last two ADs were installed. The full configuration of the Daya Bay experiment started data taking on 19 October 2012, running reliably to present. Two additional results on
neutrino oscillation were reported subsequently. With all 217 days of data acquired in the first phase, a spectral and rate analysis improved the precision of $\theta_{13}$ and measured the effective mass splitting $\Delta m^2_{ee}$ for the first time~\cite{DB_spectana}.
A new analysis with 621 days of data, including the 6-AD phase and the full 8-AD configuration, was released recently; the measured $\sin^22\theta_{13}$ has reached a precision of 6\%~\cite{DB_2015}. These results are summarized in Table~\ref{tab:sens}.

\begin{table}[!htb]
\begin{center}
\caption{Daya Bay measurements on $\sin^22\theta_{13}$ and $\Delta m^2_{ee}$. The first uncertainty of $\sin^22\theta_{13}$ in the first two rows is the statistical error and the second is the systematic one.
The unit of $\Delta m^2_{ee}$ is $10^{-3}$ eV$^2$. \label{tab:sens} }
\begin{tabular}[c]{ccccc} \hline
Release time & Data & Config & $\sin^22\theta_{13}$ & $\Delta m^2_{ee} $ \\\hline
2012/3/8~\cite{DB_discovery} & 55 days & 6 ADs & $0.092\pm0.016\pm0.005$ & - \\
2012/10/23~\cite{DB_CPC} & 139 days & 6 ADs & $0.089\pm0.010\pm0.005$ & - \\
2013/10/24~\cite{DB_spectana} & 217 days & 6 ADs & $0.090^{+0.008}_{-0.009}$ & $2.59^{+0.19}_{-0.20}$ \\
2014/6/24~\cite{DB_nH} & 217 days & 6 ADs (nH) & $0.083\pm0.018$ & - \\
2015/5/13~\cite{DB_2015} & 621 days & 6+8 ADs & $0.084\pm0.005$ & $2.42\pm0.11$ \\\hline
\end{tabular}
\end{center}
\end{table}

Daya Bay has accumulated the largest reactor antineutrino sample in the world, which enables many precision measurements. The most precise reactor antineutrino spectrum has been measured~\cite{DB_reactor}. A search for a sterile neutrino has significantly extended the exclusion area in the low-mass region of the $\sin^22\theta_{14}$-$\Delta m^2_{41}$ parameter space~\cite{DB_sterile}. Many exotic searches are ongoing. The Daya Bay experiment plans to operate until 2020. A 3\%-precision measurement on both $\sin^22\theta_{13}$ and $\Delta m^2_{ee}$ is expected.

\section{Design and Features}
\label{design}

In a reactor neutrino experiment, the sensitivity or precision in $\theta_{13}$  depends on how well the rate deficit and distortion in the energy spectrum are determined. When the exposure, defined as the product of the target mass of the far site detector in tons, the thermal power of the reactor in GW, and the live time in years, is larger than 10,000 GW-ton-yr, distortion in the energy spectrum, thus statistics, will dominate the sensitivity~\cite{huber2002}. Such exposure corresponds to about 8 years for Daya Bay. Therefore, for most cases, the rate deficit will dominate the sensitivity, and related systematic uncertainties, including the detection efficiency, target proton number, and backgrounds, should be controlled to $\lesssim 0.5$\% to measure $\sin^22\theta_{13}$ to 0.01 at 90\% confidence level.

\par
When multiple reactor cores are spread out over a large area, a single near site can only constrain the antineutrino flux from the nearby cores. In this case, the reactor-related uncertainties cannot be completely cancelled by the near-far relative measurement. Moving the near site farther away from the cores will improve the cancellation but lose sensitivity due to an increase in the oscillation effect. To obtain the best sensitivity, Daya Bay is configured with one far site for observing the maximal oscillation effect, and two near sites for determining the flux of the reactor antineutrinos from the Daya Bay and Ling Ao NPPs. The best locations of the three halls were determined with a $\chi^2$ method~\cite{sunyx}, with the projected uncertainties and estimated background at the candidate sites derived from geological survey information. For the optimal configuration, the uncertainty related to the reactors is reduced to 5\% of the uncorrelated uncertainty of a single core (0.8\%), which is totally insignificant.

\par
Experience and lessons learned in CHOOZ~\cite{chooz}, Palo Verde~\cite{paloverde}, and KamLAND~\cite{kamland} were taken into account in designing the Daya Bay detectors. Some unique features of the Daya Bay design significantly improve the detector performance; indeed, the built-in redundancy is crucial for precision measurements. Details of the Daya Bay detector design and fabrication can be found in Ref.~\cite{longdet}. In the following we will briefly describe the experimental design highlighting the concept of multiple muon taggers, multiple antineutrino detectors, a remote-controlled calibration system, photon reflectors and shields, as well as the optimization of the detector dimensions.

\par
\begin{figure}[!htb]
\begin{center}
\includegraphics[width=0.6\textwidth]{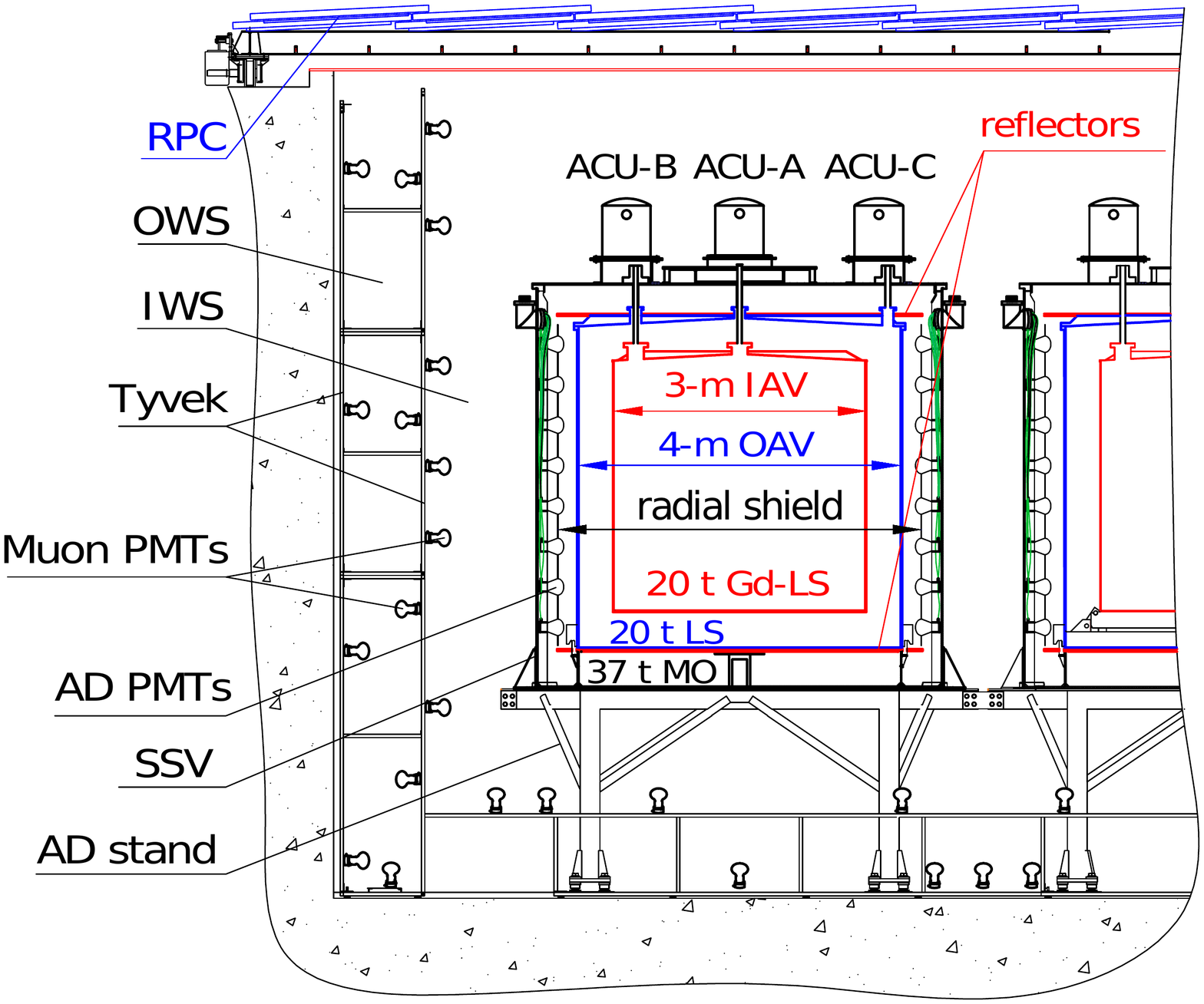}
\caption{Schematic diagram of the Daya Bay detectors. \label{fig:det}}
\end{center}
\end{figure}

Each AD has three nested cylindrical volumes separated by concentric acrylic vessels as shown in Fig.~\ref{fig:det}.  Serving as the target for the inverse beta-decay reaction, the innermost volume holds 20~t of gadolinium-doped liquid scintillator (Gd-LS) with 0.1\% Gd by weight~\cite{ding,ding2,yeh,lsproduction}. The middle volume is called the gamma catcher and is filled with 20~t of undoped liquid scintillator (LS) for detecting gamma-rays that escape the target volume.  The outer volume contains 37~t of mineral oil (MO) to provide optical homogeneity and to shield the inner volumes against radiation from the detector components.  There are 192 20-cm PMTs (Hamamatsu R5912) installed in the MO volume and around the circumference of the stainless steel vessel (SSV). The top and the bottom surfaces are not instrumented with PMTs; instead, there are two highly reflective panels. The PMTs are recessed in a 3-mm-thick black cylindrical shield located at the equator of the PMT bulb. In each hall, the ADs are submerged in a water pool that provides at least 2.5 m of water to degrade radiation from the rock. The water pool is optically divided into the inner (IWS) and outer (OWS) regions, both equipped with 20-cm PMTs to serve as water Cherenkov detectors. On the top of the water pool, there are Resistive Plate Chambers (RPCs) serving as another muon detector.

\par 
The multiple-module scheme is the most prominent feature of Daya Bay. The current generation of reactor neutrino experiments for determining $\theta_{13}$  planned to achieve a relative detector uncertainty of (0.38-0.6)\%. Such uncertainties need very careful validation. With at least two ADs in each experimental hall, comparison of performance among the ADs at the same site can actually ``measure" the relative uncertainty between them. By comparing the components of the detector responses, the relative uncertainty of two ADs was found to be 0.2\%~\cite{ad12}. The measured ratio of the antineutrino rates in two ADs was
$0.981\pm0.004~(1.019\pm0.004)$ while the expected ratio was 0.982 (1.012) for the Daya Bay (Ling Ao) near site~\cite{DB_2015}, validating the uncertainty estimation. The deviation from unity is due to slightly different baselines of the two ADs to the reactor cores. Furthermore, the uncertainties of the ADs are found to be largely uncorrelated. Therefore, the total relative detector uncertainty is statistically reduced by $1/\sqrt{N}$, where $N$ is the number of ADs at a given site.

\par 
The water pool is divided into two water Cherenkov detectors, as shown in Fig.~\ref{fig:det}. The outer one is 1 m thick and the inner one is 1.5 m. The top of the water pool is covered by RPC tiles.  The outer Cherenkov counter and the RPC play important roles in determining the fast neutron background originating from muon spallation in the surrounding rock. The efficiency of the water Cherenkov detectors for detecting muons was designed to be 95\%, and 90\% for the RPCs. The combined efficiency was aimed to be ($99.5\pm0.25$)\%. As it turned out, due to high reflectivity of the custom-made Tyvek\textsuperscript{\textregistered} composite film lining the partitions and the well-designed water purification system, the efficiency of the inner Cherenkov detector was 99.98\%~\cite{DB_muon}. Again, the multiple-detector design provides robust tagging of the cosmic-ray muons, which is essential for rejecting muon-induced background.

\par 
The geometry of the AD was optimized with extensive Monte Carlo simulation. The MO shielding is thin and the distance from the apex of the PMT to the liquid scintillator is only 20 cm. This distance is driven by the requirement of uniform detector response instead of radiation shielding. As a result, Daya Bay has a relatively high singles rate and accidental coincidence to maximize the target mass. Since the accidental-coincidence background can be determined accurately to high precision, it has negligible impact to the sensitivity or precision.

\par 
To obtain good light yield with fewer number of PMTs, reflective panels are used.
A specular reflective film with reflectivity $>98\%$ in the scintillation light region, ESR (3M\textsuperscript{\textregistered}), is sandwiched between two 1-cm-thick acrylic panels 4.5 m in diameter. The space between the two acrylic panels is evacuated when bonding together on the edge. The sandwich structure is maintained by vacuum pressure with the least mechanical connections between two panels to keep the optical surface intact. A reflector is put on the top of the gamma catcher and another at the bottom. Such a design reduces the number of PMTs by about one half while achieving almost the same energy and vertex resolution. Furthermore, the mechanical structure of the AD is simplified with the adoption of reflective panels, enabling transportable ADs.

\par 
Three water-proof automated calibration units (ACUs) are mounted on the top of each AD. Each ACU is equipped with a LED, a $^{68}$Ge source, and a composite source of $^{241}$Am-$^{13}$C and $^{60}$Co. Deployment of the source into the liquid scintillator, one at a time, is controlled remotely. With the ACUs, the AD can be fully submerged in water without a chimney for calibration. Therefore, Daya Bay does not experience backgrounds coming from external radioactivity or the Michel electrons from decays of stopped muon.

\par 
The inner wall of the stainless steel tank is painted black with a fluor-carbon paint. To reduce the stray light reflected from the PMT glass, cable, and other components, another light shield made of black ABS is installed at the equator of the PMTs. This design has an unforeseen advantage of suppressing an instrumental background, the PMT flasher events, which appeared in many neutrino experiments using PMTs but become difficult to reject for relative small detectors. The electronic components or connections on the base of the Hamamatsu PMT may occasionally discharge and produce a flash of light. The detected energy of these faked events ranges from sub-MeV to a hundred MeV in Daya Bay. Although only a small fraction of the PMTs spontaneously discharge infrequently, these flasher events significantly increase the accidental-coincidence background. With the black shield, the flasher events always appear as a characteristic PMT-hit pattern; thus they can be easily identified and rejected~\cite{DB_CPC}. Without this unique PMT-hit pattern, we would have to either turn off the flashing PMTs, or bear a larger uncertainty in the selection efficiency and a larger accidental-coincidence background.

\section{Signal and Background}
\label{signal}
\par
The reactor antineutrinos are detected via the inverse $\beta$-decay (IBD) reaction, $\overline{\nu}_e + p \to e^+ + n$, in the Gd-LS\@. The coincidence of the prompt scintillation from the $e^+$ and the delayed neutron capture on Gd provides a distinctive signature. The positron carries almost all of the kinetic energy of the antineutrino, thus the positron energy deposited in the liquid scintillator is highly correlated with the antineutrino energy. The neutron thermalizes before being captured on either a proton or a gadolinium nucleus with a mean capture time of $\sim$30 $\mu$s in Gd-LS or $\sim$200 $\mu$s in normal LS. When a neutron is captured on Gd, it releases several gamma-rays with a total energy of $\sim$8~MeV, and is thus easily distinguished from the background coming from natural radioactivity. The capture on H suffers from a larger background but provides an independent measurement and can improve the precision of the $\theta_{13}$ measurement.

\par
The ADs are calibrated with sources in the ACUs weekly, and with spallation products, IBDs, and natural radiation from materials inside the detectors. Two independent calibration algorithms are utilized. The energy scale is determined using the $^{60}$Co or Am-C neutron source in the ACUs, or spallation neutrons. The uncertainty in the energy scale is determined by comparing more than 10 known references in the 8 ADs and by studying their stabilities over time. The energy scale uncertainty has reduced from 0.5\% reported in the initial publications~\cite{ad12,DB_discovery} to 0.2\% in the latest~\cite{DB_2015}. The reduction comes from the improvements in the correction of position and time dependence.

\par
Nonlinearity in the energy response of an AD originates from two dominant sources: particle-dependent nonlinear light yield of the scintillator and charge-dependent nonlinearity in the PMT readout electronics. Each effect is at the level of 10\%. We have constructed a semi-empirical model that predicts the reconstructed energy for a particle assuming a specific energy deposited in the scintillator. The model contains four parameters: Birks' constant, the relative contribution to the total light yield from Cherenkov radiation, and the amplitude and scale of an exponential correction describing the non-linear electronics response. This exponential form of the electronics response is motivated by Monte Carlo (MC) and data; it has been confirmed with an independent FADC measurement. Besides the calibration references used in the energy scale studies, a broad set of calibration sources were deployed into the two ADs of EH1 using the ACUs and a manual calibration system~\cite{huanghx} during the shutdown in the summer of 2012. The energy nonlinearity, i.e. the absolute energy scale, is determined to $<1$\% above 2 MeV.

\par
To select reactor antineutrino events, the PMT flasher background is rejected first. The prompt and delayed signals are required to be 0.7-12 MeV and 6-12 MeV, respectively. The temporal separation between a pair of prompt and delayed signals should be within 1-200 $\mu$s. To reject cosmogenic backgrounds, the delayed signal is required to be 600 $\mu$s, 1000 $\mu$s, or 1 s later than a muon, depending on the deposited energy of the muon. Finally, no other signal should occur in 200 $\mu$s before the prompt signal and after the delayed signal. Two independent algorithms are developed following these criteria with minor differences. The selected samples differ by less than 10\%, mostly due to the different energy calibration used.

\par
A detailed treatment of the absolute and relative efficiencies was reported in Refs.~\cite{ad12,DB_CPC}. The uncertainties of the absolute efficiencies are correlated among the ADs and are thus negligible in oscillation analyses. The determination of all relative uncertainties is data-driven. The dominant ones come from the energy calibration and the neutron capture fraction on Gd, both at the 0.1\% level. The total relative uncertainty is conservatively estimated to be 0.2\%, uncorrelated among ADs.

\par
Five kinds of background are considered in Daya Bay. The accidental background is the largest one but contributes negligible to the total systematic uncertainty. The most serious background is the cosmogenic $^9$Li/$^8$He, which contributes an uncertainty of $\sim0.2$\%. The remaining three kinds of background have an uncertainty of $\sim 0.01$\%.

\par
The accidental background, from accidental correlation of two unrelated signals, is determined by measuring the rate of both prompt- and delayed-like signals, and then estimating the probability that two signals randomly satisfy the time coincidence for the IBD selection.

\par
The $^9$Li/$^8$He background comes from the $\beta$-n decay of $^9$Li/$^8$He produced by muons in the ADs. The rate is evaluated from the distribution of the time since the last muon using the known decay times for these isotopes. A 50\% systematic uncertainty is assigned to account for the extrapolation to zero deposited muon energy.

\par
An energetic neutron entering an AD can form a fast-neutron background by recoiling off
a proton before being captured. It is estimated by extrapolating the prompt energy spectrum into the IBD energy region, or by studying the muon-tagged fast neutron sample.

\par
The $^{13}$C($\alpha$,n)$^{16}$O background was determined using MC after estimating the amount of $^{238}$U, $^{232}$Th, $^{227}$Ac, and $^{210}$Po in the \mbox{Gd-LS} from their cascade decays, or by fitting their $\alpha$-particle energy peaks in the data.

\par
A neutron emitted from the \mbox{Am-C} neutron source in an ACU could generate a gamma-ray via inelastic scattering in the stainless steel vessel before subsequently being captured on Fe/Cr/Mn/Ni. An IBD is mimicked if both gamma-rays from the scattering and capture processes enter the scintillator. This correlated background is estimated using MC\@.  The normalization is constrained by the measured rate of single delayed-like candidates from this source.

\section{Oscillation analyses}
\label{oscillation}

Early results of Daya Bay were based on rate analysis when the statistics were low. The rate deficit at the far site was $\sim6$\% compared to the prediction based on a weighted combination of two near site measurements. The value of $\sin^22\theta_{13}$ is determined with a $\chi^2$ constructed with pull terms accounting for the correlation of the systematic errors,
\begin{eqnarray}  \label{eqn:chi2}
 \chi^2 &=&
 \sum_{d=1}^{6}
 \frac{\left[M_d-T_d\left(1+  \varepsilon
 + \sum_r\omega_r^d\alpha_r
 + \varepsilon_d\right) +\eta_d\right]^2}
 {M_d+B_d}  \nonumber \\
 &+&
 \sum_r\frac{\alpha_r^2}{\sigma_r^2}
 + \sum_{d=1}^{6} \left(
 \frac{\varepsilon_d^2}{\sigma_d^2}
 + \frac{\eta_d^2}{\sigma_{B}^2}
 \right)
 \,,
\end{eqnarray}
where $M_d$ are the measured number of IBD events of the $d$-th AD with its backgrounds subtracted, $B_d$ is the corresponding background, $T_d$ is the prediction from antineutrino flux, including MC corrections for energy responses and neutrino oscillations, $\omega_r^d$ is the fraction of IBD contribution of the $r$-th reactor to the $d$-th AD determined by the baselines and antineutrino fluxes. The uncorrelated reactor uncertainty is $\sigma_r$ (0.8\%). The parameter $\sigma_d$ (0.2\%) is the uncorrelated detection uncertainty. The parameter $\sigma_{B}$ ($\sim0.2$\%) is the quadratic sum of the background uncertainties, which are site dependent. The corresponding pull parameters are ($\alpha_r, \varepsilon_d, \eta_d$). The absolute normalization $\varepsilon$, which absorbs the detector- and reactor-related correlated uncertainties, is a free parameter determined from the fit to the data. While keeping $\varepsilon$ free, the reactor antineutrino flux is determined by the near site measurements. The model-dependent reactor flux prediction enters the fit only at secondary order.

\par
With increased statistics, the latest Daya Bay analyses are based on rate and shape analysis. While rate deficit still dominates the $\theta_{13}$ sensitivity, the spectral information starts to contribute. To take advantage of the spectral information, an analysis with a similar $\chi^2$ expression as defined in Eq.~\ref{eqn:chi2} but with energy bins and relevant uncertainties is used. Additional inputs to the fit include the background shape uncertainties and energy nonlinearity model described in Sec.~\ref{signal}. Besides improving the precision of the $\sin^22\theta_{13}$, the effective mass splitting $\Delta m^2_{ee}$ has been measured for the first time~\cite{DB_spectana}.

\par
\begin{figure}[!htb]
\begin{center}
\includegraphics[width=0.6\textwidth]{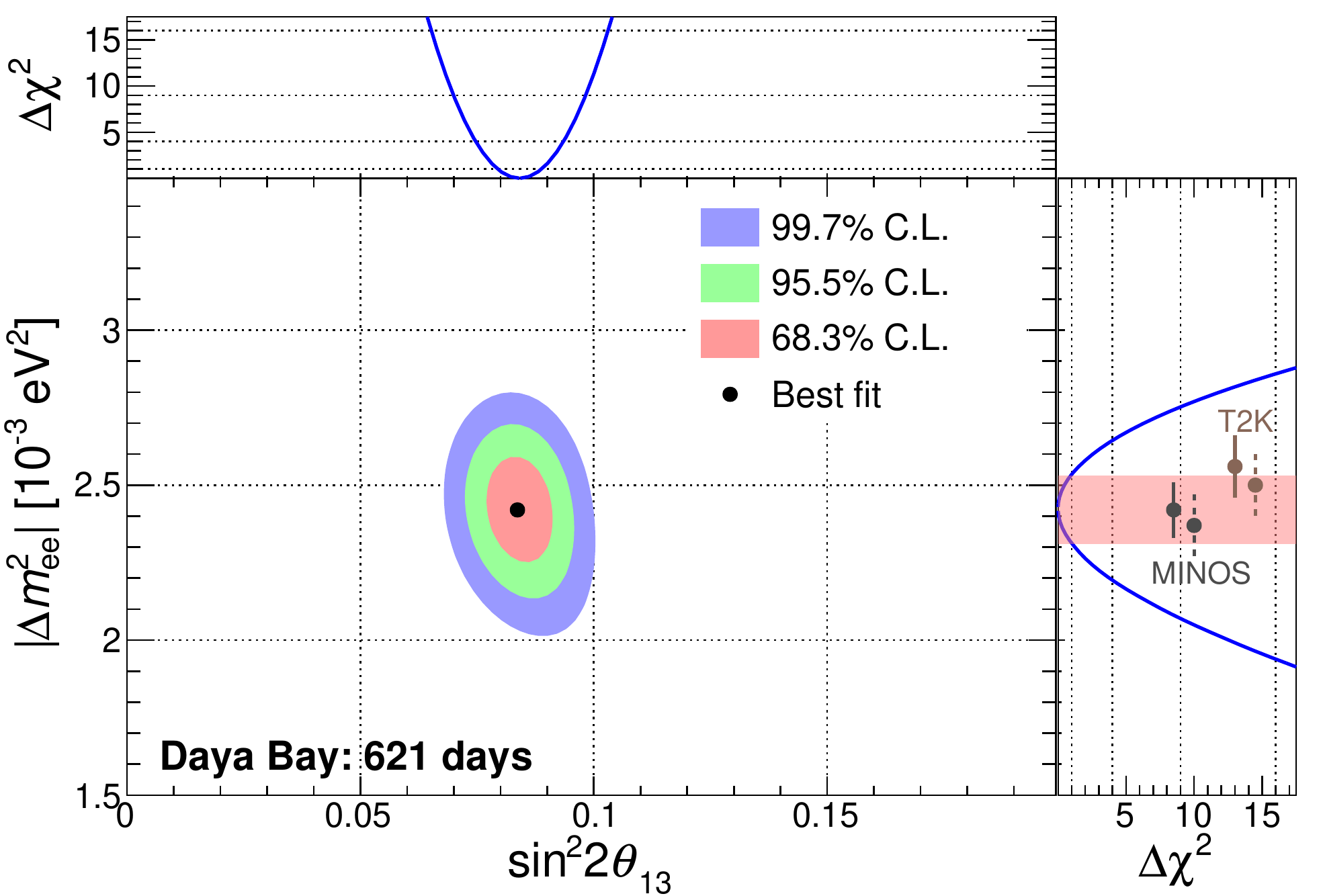}
\caption{Error contours corresponding to the $68.3\%$, $95.5\%$ and $99.7\%$ confidence levels in the $|\Delta m^{2}_{\mathrm{ee}}|$-$\sin^{2}2\theta_{13}$ plane. The contours were
obtained with a least-squares fit given in Eq.~\ref{eq:chi22} using the measured $\overline{\nu}_e$ rates and energy spectra at the near and far sites. The adjoining panels show the dependence
of $\Delta\chi^{2}$ on $\sin^{2}2\theta_{13}$ (top) and $|\Delta m^2_{\mathrm{ee}}|$ (right). Figure adapted from~\cite{DB_2015}. \label{fig:contours}}
\end{center}
\end{figure}

Another approach in extracting the oscillation parameters is to construct a $\chi^2$ expression using a covariance matrix
\begin{equation}
  \label{eq:chi22}
  \chi^{2} = \sum_{i,j}(N_j^{\mathrm{f}} - w_j \cdot N_j^{\mathrm{n}}) (V^{-1})_{ij} (N_i^{\mathrm{f}} - w_i \cdot N_i^{\mathrm{n}}),
\end{equation}
where $N_i$ is the observed number of events after background
subtraction in the $i$-th bin of reconstructed positron energy. The superscript $f~(n)$ denotes a far~(near) detector. The symbol $V$ represents a covariance matrix that includes
known systematic and statistical uncertainties.  The quantity $w_i$ is
a weight that accounts for the differences between the near- and far-site measurements. In this method, the flux and spectrum of the antineutrinos at the reactor play a negligible role~\cite{DB_2015}. The results from the fit shown in the $\sin^22\theta_{13}$-$\Delta m^2_{ee}$ plane are depicted in Fig.~\ref{fig:contours}.

\par
Daya Bay has also accumulated an IBD sample with neutron capture on hydrogen (nH) which has comparable statistics to that of neutron capture on gadolinium (nGd). Since the delayed signal of 2.2 MeV falls into the energy region totally dominated by natural radiation, the coincidence background is huge. By requiring the reconstructed distance between the positron and delayed-neutron vertices be $<50$ cm, 98\% of this background can be rejected while losing 25\% of the signal. A spectral subtraction is further needed to remove the accidental backgrounds. An analysis using the 217-day data set yielded $\sin^22\theta_{13}=0.080\pm0.018$. Since the IBD sample and the systematic uncertainties are largely independent from the nGd analysis, the nH analysis provides an independent measurement of $\theta_{13}$. The correlation between the nH and nGd analyses is evaluated to be 0.05. When the nH and nGd analyses of this data set are combined, the $\sin^22\theta_{13}$ sensitivity is improved by 8\%~\cite{DB_nH}.

\section{Reactor Antineutrino Spectrum and Exotic searches}
\label{reactor}
\par
Although the oscillation analyses of Daya Bay have negligible dependence on the external prediction of the reactor antineutrino flux and spectrum, the knowledge is a crucial factor for many reactor experiments. In early studies, the estimation relied on calculations or other indirect means, such as the $\beta$ spectrum measurements made on reactor fuels, based on the understanding of the complex fission processes in the reactor core. These methods have rather strong dependence on theoretical models. Daya Bay has accumulated the world's largest sample of reactor antineutrinos, at a rate of $\sim 1$ million per year. A direct measurement will provide the most precise and model independent  reactor antineutrino spectrum.

\par
In reactor cores, antineutrinos are emitted from four primary fuel isotopes: $^{235}$U, $^{238}$U, $^{239}$Pu, and $^{241}$Pu. Each fission releases about 200 MeV energy ((0.2-0.5)\% uncertainty). Fission rates can be estimated with the thermal power measurement (0.5\% uncertainty) and core simulation of the evolution of the fuel composition (0.6\% uncertainty with the constraint of the total thermal power)~\cite{cjflux}. The most uncertain part is the rate and spectrum of antineutrinos emitted from each fission of the four isotopes. By fitting the measured $\beta$ spectrum of the $^{235}$U, $^{239}$Pu, and $^{241}$Pu fuel~\cite{illschr,illvonf,illhahn} with hypothesized virtual $\beta$ decays, plus theoretical calculation for $^{238}$U, two models, the ILL-Vogel and the Huber-Mueller models, have been developed~\cite{vogel238,mueller,huber} with an uncertainty of (2-3)\%.

\par
The latest Daya Bay measurement on the absolute reactor antineutrino rate is $\sigma_f=(5.92\pm0.14)\times 10^{-43} {\rm cm}^2\, {\rm fission}^{-1}$, where the dominant uncertainty comes from the absolute efficiency (2.1\%)~\cite{DB_reactor}. Comparing to the Huber-Mueller model, there is a $\sim6$\% deficit, consistent with the past 19 short baseline ($<100$ m) measurements.

\par
With the nonlinearity model described in Sec.~\ref{signal}, Daya Bay has measured the prompt-energy spectrum with a precision ranging from 1.0\% at 3.5 MeV to 6.7\% at 7 MeV. An excess of about 10\% at $\sim 5$ MeV compared with expectations is observed, leading to a discrepancy of up to $4~\sigma$.

\par
\begin{figure}[!htb]
\begin{center}
\includegraphics[width=0.6\textwidth]{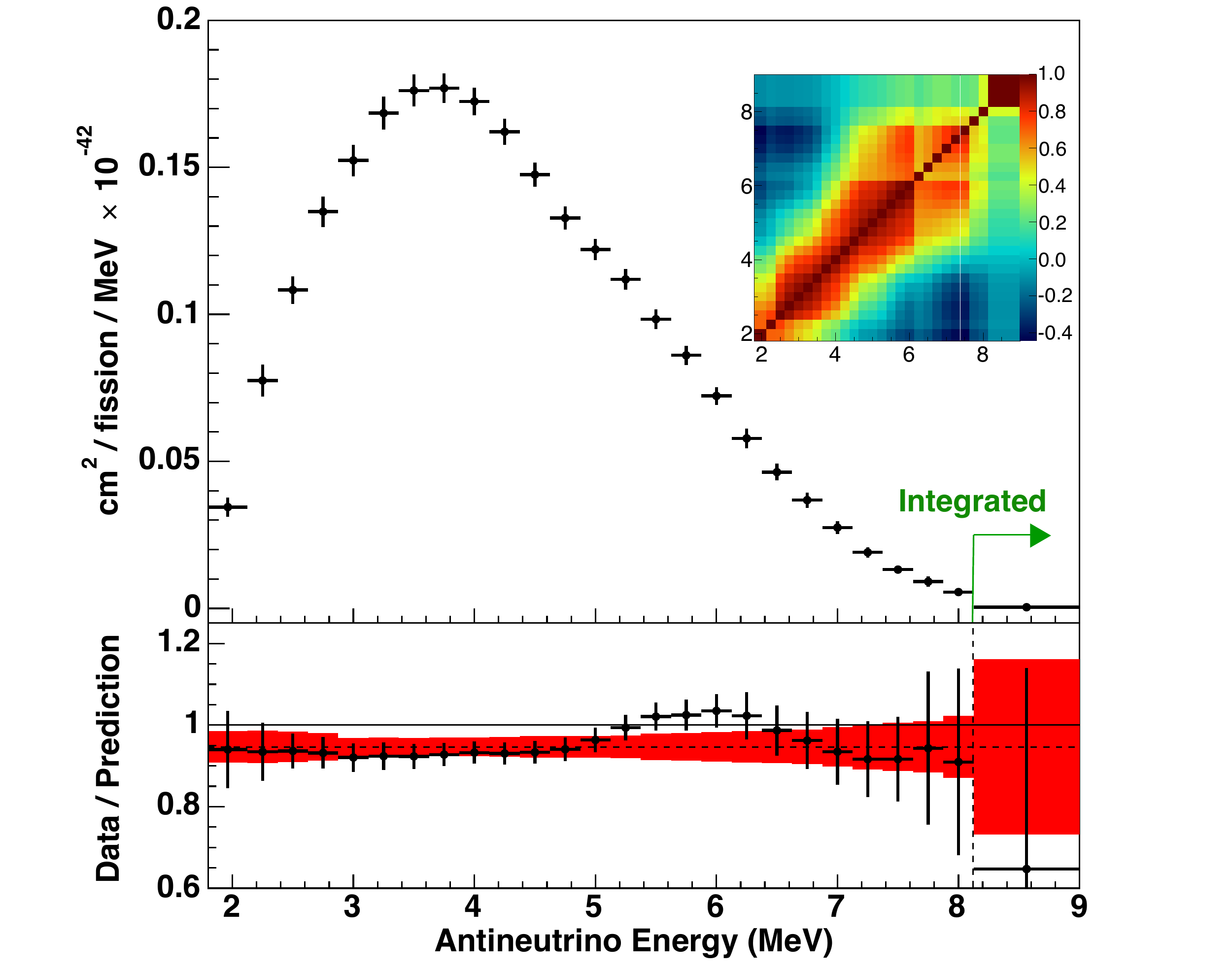}
\caption{Top panel: The extracted reactor antineutrino spectrum and its correlation matrix.
Bottom panel: Ratio of the extracted reactor antineutrino spectrum to the Huber+Mueller prediction. Figure adapted from~\cite{DB_reactor}.
\label{fig:generic}}
\end{center}
\end{figure}

Such a deviation shows the importance of the direct measurement of the reactor antineutrino spectrum, particularly for next-generation reactor experiments such as JUNO~\cite{JUNO}, and may indicate the need to revisit the models underlying the calculations. The prompt-energy spectrum is unfolded, i.e. removing the detector response effect, to an antineutrino spectrum, as shown in Fig.~\ref{fig:generic}.

\par
Daya Bay will improve the measurement with much larger statistics and better energy nonlinearity. Recently a Flash Analog-to-Digital Converter (FADC) system has been installed on one AD to pin down the nonlinearity from the electronics. Another calibration and systematic study campaign is being planned. We expect to measure the reactor antineutrino spectrum to 1\% precision in a large energy range, and improve the precision especially for the very high- and very low-energy regions which are outside the ranges of current models.

\par
The high precision antineutrino spectrum is also excellent for light sterile neutrino searches. If light sterile neutrinos mix with the three active neutrinos, their presence could be detected by looking for the fast oscillatory behavior in the spectrum. Daya Bay has significantly extended the exclusion area in $10^{-3} {\rm eV}^2 \lesssim |\Delta m^2_{41}|\lesssim 0.1 {\rm eV}^2$~\cite{DB_sterile}. Further improvements with increasing statistics are expected.

\par
Besides the sterile neutrino studies, more exotic searches are in progress, e.g. for the non-standard interaction, decoherence effect, mass-varying neutrino, Lorentz-violation and CPT violation, etc.

\section{Summary and Prospect}
\label{summary}

\par
\begin{figure}[!htb]
\begin{center}
\includegraphics[width=0.6\textwidth]{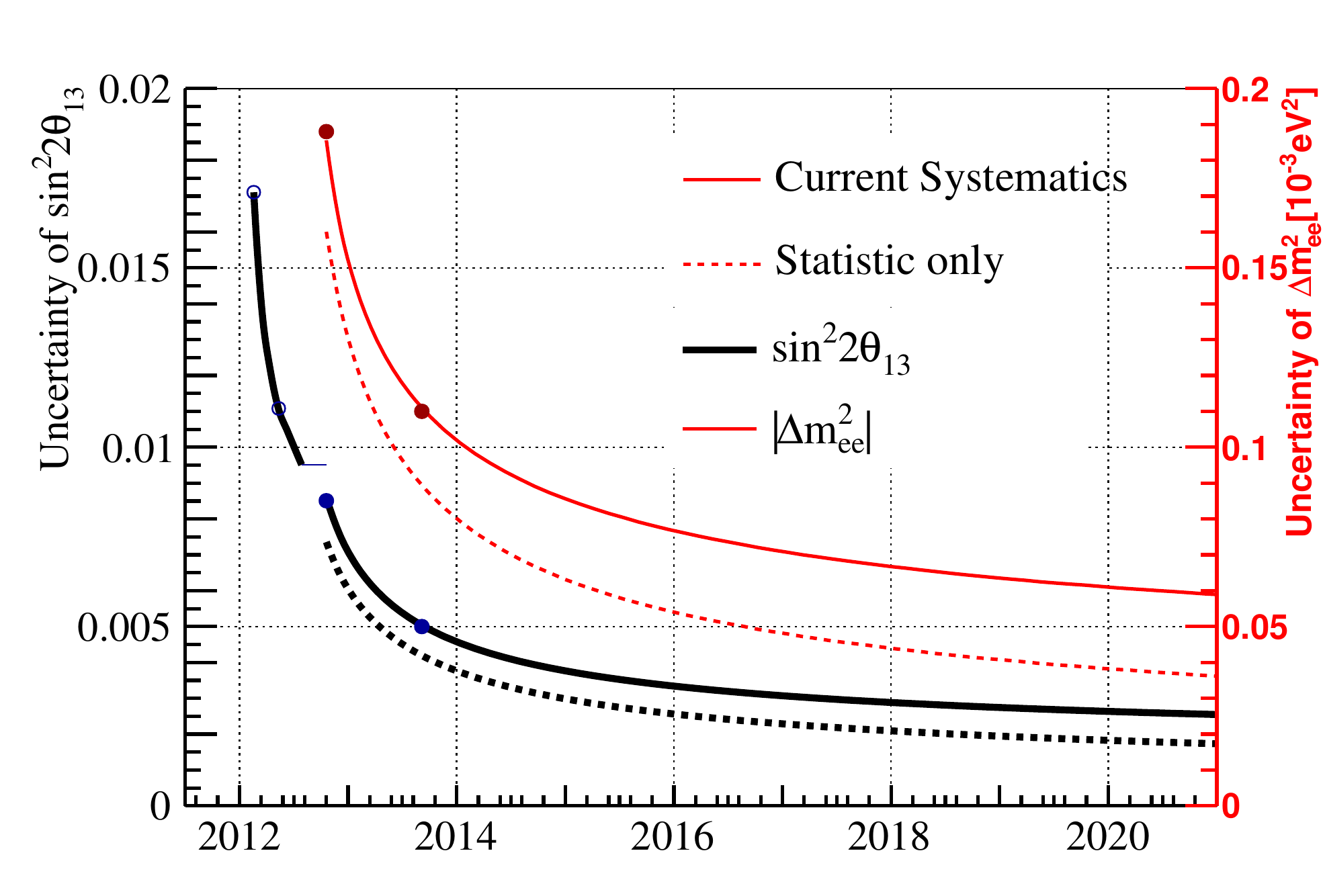}
\caption{Projected precision of $\sin^22\theta_{13}$ (black thick lines) and $\Delta m^2_{ee}$ (red thin lines) in Daya Bay, where the solid lines present the precision estimated with current systematics and the dashed lines show the statistical limit with zero systematic uncertainty. The points on the curves show the precision of published Daya Bay results.\label{fig:future}}
\end{center}
\end{figure}

With an almost ideal experimental site and unique design, the Daya Bay experiment has excellent capability for high precision measurements of reactor antineutrinos. We have reviewed our design experience, which may help future reactor neutrino experiments. The measurements on $\theta_{13}$ and effective mass splitting are reviewed. Current precision on $\sin^22\theta_{13}$ and $|\Delta m^2_{\mathrm{ee}}|$ are 6\% and 4.5\%, respectively. The projected precisions are shown in Fig.~\ref{fig:future}. The Daya Bay experiment is expected to operate until 2020; by then, the precision is $\sim3$\% for both $\sin^22\theta_{13}$ and $|\Delta m^2_{\mathrm{ee}}|$. Daya Bay has also obtained the most precise reactor antineutrino spectrum, which will
be very valuable for designing the next-generation reactor neutrino experiments that depend on
this input, such as JUNO.

\section*{Acknowledgement}

We would like to thank Jie Zhao for preparing the Fig.~\ref{fig:future}. J.C.~is partially supported by the National Natural Science Foundation of China (11225525) and K.B.L.~is partially supported by the U.S. Department of Energy, OHEP DE-AC02-05CH11231.

\end{document}